\documentclass[11pt,a4paper]{article}

\usepackage{epsfig,amsmath,amssymb,cite}

\begin{document}

\title{Strong deflection lensing by charged black holes \\ in scalar--tensor gravity}
\author{Ernesto F. Eiroa$^{1,2,}$\thanks{e-mail: eiroa@iafe.uba.ar}, Carlos M. Sendra$^{1,2,}$\thanks{e-mail: cmsendra@iafe.uba.ar} \\
{\small $^1$ Instituto de Astronom\'{\i}a y F\'{\i}sica del Espacio (IAFE, CONICET-UBA),} \\
{\small Casilla de Correo 67, Sucursal 28, 1428 Buenos Aires, Argentina}\\
{\small $^2$ Departamento de F\'{\i}sica, Facultad de Ciencias Exactas y
Naturales, Universidad de Buenos Aires,} \\
{\small Ciudad Universitaria Pabell\'on I, 1428 Buenos Aires, Argentina} }

\maketitle

\date{}

\begin{abstract}

We examine a class of charged black holes in scalar--tensor gravity as gravitational lenses. We find the deflection angle in the strong deflection limit, from which we obtain the positions and the magnifications of the relativistic images. We compare our results with those corresponding to the Reissner--Norstr\"om spacetime and we analyze the observational aspects in the case of the Galactic supermassive black hole.

\end{abstract}



\section{Introduction}

The discovery of supermassive black holes at the center of galaxies, specially the one corresponding to SgrA* in the  Milky Way \cite{guillessen}, has led to a growing interest in the study of strong deflection gravitational lensing. Within this context, an important feature is that the observation of optical effects due to the supermassive Galactic black hole, including direct imaging, seems to be possible in the near future \cite{zakharov,johannsen,sga}. An astrophysical object with a photon sphere makes light rays passing close to it to have a large deviation, resulting in two infinite sets of the denominated relativistic images \cite{virbha1}. In this case, an analytical treatment can be performed by using the strong deflection limit, which was introduced for the Schwarzschild geometry \cite{darwin-otros}, extended to the Reissner--Nordstr\"om spacetime \cite{eiroto}, and to any spherically symmetric object \cite{bozza}. Using this method, which consists in a logarithmic approximation of the deflection angle for light rays deflecting close to the photon sphere, it is possible to obtain the positions, the magnifications, and the time delays of the relativistic images. Many works considering strong deflection lenses with spherical symmetry, most of them analytical and the others numerical, can be found in the literature \cite{nakedsing1,nakedsing2,lenseq,virbha2,alternative,bwlens}. The lensing effects of rotating black holes were also analyzed in several articles \cite{rotbh1,rotbh2}; a related interesting aspect is that the apparent shapes (or shadows) of rotating black holes have a deformation produced by the spin \cite{rotbh2,shadow1,shadow2}. For recent reviews about strong deflection lensing see \cite{reviewlens}.

The usual explanation for the accelerated expansion \cite{snova} of the Universe is that it is filled with a negative pressure fluid called dark energy \cite{de} which represents about 70 \% of the total, while the other 30 \% corresponds to visible and dark matter. The simplest equation of state for the main component is the linear expression $p=w\rho$, between the pressure  $p$ and the energy density $\rho$; depending on the values of $w$ the fluid receives different names: quintessence ($w>-1$), cosmological constant ($w=-1$), and phantom energy ($w<-1$). Dark energy can be modeled by a self-interacting scalar field with a potential \cite{de}. In this context, dilaton and phantom field solutions with spherical symmetry corresponding to black holes, wormholes, and black universes were found by several authors \cite{gibbfbdm}; while black holes in scalar--tensor gravity were analyzed in the works \cite{bronche,sofa}. Phantom black holes \cite{phalens} and black holes in Brans--Dicke theory \cite{sdlbd} were recently studied as gravitational lenses. 

In this article we investigate as gravitational lenses a class of charged black holes in scalar--tensor gravity introduced in Ref. \cite{bronche}. In the framework of General Relativity with a minimally coupled scalar $\phi $ and electromagnetic $F_{\mu \nu }$ fields as sources, we consider in the Einstein frame the Lagrangian (in units such that $8\pi G=c=1$) given by
\begin{equation}
L = \frac{1}{2} [R + g^{\mu \nu} \phi_{,\mu }\phi_{,\nu}- 2 V(\phi)-F^{\mu \nu }F_{\mu \nu }].
\label{lagrangian}
\end{equation}
with $R$ the Ricci scalar and $V(\phi)$ the scalar field potential. The Einstein-scalar equations resulting from this Lagrangian admit a static and spherically symmetric solution in the form
\begin{equation}
ds^{2}=A(\rho )dt^{2}-B(\rho )dr^{2}-C(\rho )(d\theta^{2}+\sin^{2}\theta d\phi^{2}),
\label{m1}
\end{equation}
with metric functions
\begin{equation}
A(\rho )= A_0 r^2 + 1 + 3M\left[-\frac{\rho}{b^2}
+\frac{r^2}{2b^3} \ln\frac{\rho+b}{\rho-b}\right]
-\frac{Q^2}{b^4} \left[b^2-b\rho\ln\frac{\rho+b}
{\rho-b}+\frac{r^2}{4} \ln^2\frac{\rho+b}{\rho-b} \right],
\label{m1a}
\nonumber
\end{equation}
\begin{equation}
B(\rho )=\frac{1}{A(\rho )},  \qquad  C(\rho )=r^2(\rho ),
\label{m1b}
\nonumber
\end{equation}
and the scalar field
\begin{equation}
\phi(\rho) = \phi_0 + \frac{\sqrt{2}}{2}\ln\frac{\rho+b}{\rho-b},
\end{equation}
\begin{eqnarray}
V(\rho) &=& - \frac{A_0 (3\rho^2 - b^2)}{r^2} + \frac{9 M \rho r^2 + Q^2 (3\rho^2 -2b^2)}{b^2 r^4} - \frac{3M (3\rho^2-b^2) + 6 Q^2\rho}{2b^3 r^2} \ln \frac{\rho + b}{\rho - b}\nonumber \\
&& + \frac{Q^2 (3\rho^2-b^2)}{4b^4 r^2}\ln^2 \frac{\rho+b}{\rho-b},
\end{eqnarray}
where $r^2(\rho )=\rho ^2 -b^2$, $b$ is an arbitrary constant, and $M$ is the mass; $A_0$ and $\phi _0$ are integration constants. The corresponding Maxwell fields are radial electric $F_{01}F^{10}=Q_e^2/r^4$ and magnetic $F_{23}F^{23}=Q_m^2/r^4$; then $Q^2=Q_e^2+Q_m^2$ is the square of the electromagnetic charge. The parameter $b$ represents a characteristic length associated to the scalar field. The radial coordinate satisfies the inequality $\rho \ge |b|$; the value $\rho = |b|$ corresponds to the singularity. We adopt $A_0=0$, so the metric is asymptotically flat for $\rho \to \infty$; in this case the metric is approximately Reissner--Nordstr\"om for large $\rho$.

The paper is organized as follows. In Sec. \ref{defang}, we review the main physical properties of the spacetime, we introduce the lens equation, and we obtain the exact expression for the deflection angle. In Sec. \ref{sdl}, we find the strong deflection limit, from which we calculate the positions and magnifications of the relativistic images. In Sec. \ref{obs} we find the observables and we analyze the observational prospects for the case of SgrA*. Finally, in Sec. \ref{summary}, we summarize the results obtained.

\section{Deflection angle}\label{defang}

\begin{figure}[t!]
\begin{center}
    \includegraphics[scale=0.9,clip=true,angle=0]{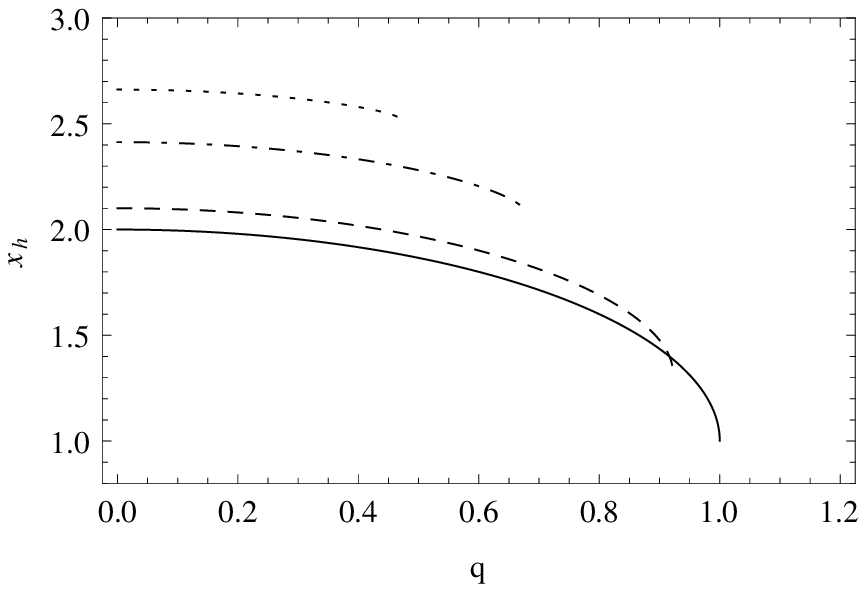}
    \includegraphics[scale=0.9,clip=true,angle=0]{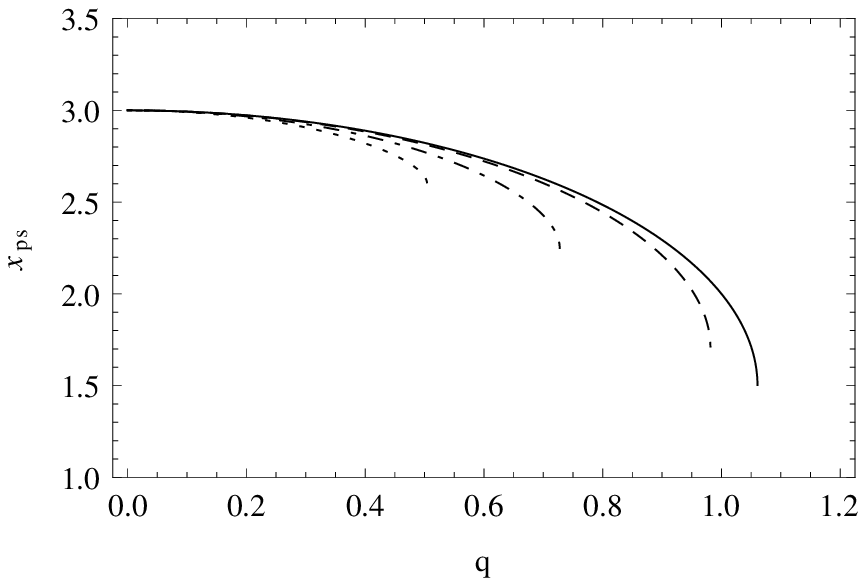}
    \caption{Adimensionalized radius of the horizon $x_h$ and of the photon sphere $x_{ps}$ as functions of the adimensionalized charge $q$ for some representative values of the adimensionalized parameter $\tilde{b}$: $1$ (dashed line), $2$ (dash-dotted line) and $2.5$ (dotted line); the plots corresponding to the Reissner--Nordstr\"om spacetime are also shown (solid line).}
    \label{hpsfig}
\end{center}
\end{figure}

We start by adopting the asymptotic flatness condition $A_0=0$ and also adimensionalizing the metric (\ref{m1}) in terms of the mass, introducing the new radial and time coordinates
\begin{equation}
x=\frac{\rho}{M}, \qquad  \tilde{t}=\frac{t}{M},
\end{equation}
and the parameters
\begin{equation}
q=\frac{Q}{M}, \qquad  \tilde{b}=\frac{b}{M};
\end{equation}
so the line element has the form
\begin{equation}
ds^2=A(x)d\tilde{t}^2-B(x)dx^2-C(x)(d\theta^2+\sin^2\theta d\phi^2),
\label{m2}
\end{equation}
where
\begin{equation}
A(x)=1+3\left[-\frac{x}{\tilde{b}^2}+\frac{(x^2-\tilde{b}^2)}{2\tilde{b}^3}\ln\frac{x+\tilde{b}}{x-\tilde{b}}\right]-\frac{q^2}{\tilde{b}^4}\left[\tilde{b}^2-\tilde{b}x\ln\frac{x+\tilde{b}}{x-\tilde{b}}+\frac{x^2-\tilde{b}^2}{4}\ln^2\frac{x+\tilde{b}}{x-\tilde{b}}\right],
\label{m2a} \nonumber
\end{equation}
\begin{equation}
B(x)=\frac{1}{A(x)}, \qquad  C(x)=x^2-\tilde{b}^2.
\label{m2b} \nonumber
\end{equation}
We can take $\tilde{b}\ge 0$ and $q\ge 0$ without losing generality because the metric is invariant under the transformations $\tilde{b} \longleftrightarrow -\tilde{b}$ and $q \longleftrightarrow -q$. The radius of the event horizon $x_h$, obtained numerically for each $\tilde{b}\neq 0$ as the largest value of $x$ satisfying the condition $A(x)=0$, is a decreasing function of the charge $q$. For a given $\tilde{b}$, there is a value of charge which corresponds to the extremal black hole; for a larger $q$ there is a naked singularity. In the limit $\tilde{b}\rightarrow 0$ the geometry (\ref{m2}) reduces to the Reissner--Nordstr\"om spacetime, with metric functions $A(x)=B(x)^{-1}=1-2x^{-1}+q^2x^{-2}$ and $C(x)=x^2$, for which the event horizon radius is given by $x_h=1+\sqrt{1-q^2}$. As $\tilde{b}$ increases, the range of values of $q$ for which there is a horizon, becomes smaller, as shown in Fig. \ref{hpsfig}. The adimensionalized radius $x_{ps}$ of the photon sphere is given by the largest positive solution of the equation
\begin{equation}
\frac{A'(x)}{A(x)}=\frac{C'(x)}{C(x)},
\label{xps}
\end{equation}
where the prime represents the derivative with respect to $x$. Replacing the metric functions and simplifying, this equation takes the form
\begin{equation}
 3-x-\frac{q^2}{\tilde{b}}\ln \frac{x+\tilde{b}}{x-\tilde{b}}=0,
\end{equation}
which can be solved numerically for the different values of the parameter $\tilde{b}\neq 0$ and the charge $q$ (except in the trivial case $q=0$ which gives $x_{ps}=3$ for all $\tilde{b}$). In the case of the Reissner--Nordstr\"om geometry we have $x_{ps}=(3+\sqrt{9-8q^2})/2$. For a given value of $\tilde{b}$, from Fig. \ref{hpsfig} we see that $x_{ps}$ decreases as the charge increases; there is a small range of $q$ for which there is a naked singularity surrounded by a photon sphere.

The exact deflection angle $\alpha$ for a photon coming from infinity, is related to the adimensionalized closest approach distance $x_0$, by the expression \cite{nakedsing1,weinberg}
\begin{equation}
\alpha(x_0)=I(x_0)-\pi,
\label{alpha1}
\end{equation}
where the integral
\begin{equation}
I(x_0)=\int^{\infty}_{x_0}\frac{2\sqrt{B(x)}dx}{\sqrt{C(x)}\sqrt{A(x_0)C(x)\left[ A(x)C(x_0)\right]^{-1}-1}},
\label{i0}
\end{equation}
is a monotonic decreasing function of $x_0$. For photons passing close to the photon sphere, the deflection angle is large and diverges in the limit $x_0\rightarrow x_{ps}$. On the other hand, when $x_0\rightarrow\infty$ we have that $\alpha\rightarrow 0$. In a gravitational lensing scenario, the deflection angle is related to the angular position of the source ($\beta$) and the images ($\theta$) by the lens equation. For the object acting as a lens ($l$) placed between a point source of light ($s$) and an observer ($o$), both situated in the flat region of spacetime, the lens equation has the form \cite{lenseq}
\begin{equation}
\tan \beta =\frac{d_{ol}\sin \theta - d_{ls} \sin (\alpha -\theta)}{d_{os} \cos (\alpha -\theta)} ,
\label{pm1}
\end{equation}
where  $d_{os}$, $d_{ol}$, and $d_{ls}$, are the observer--source, observer--lens, and lens--source adimensionalized angular diameter distances, respectively. To obtain the positions of the images $\theta$, for a given source position  $\beta$ one has to calculate the deflection angle $\alpha $ and invert the lens equation (\ref{pm1}). For an analytical treatment, some approximations are necessary for both the deflection angle and the lens equation. As we are interested in the lensing phenomena characteristic of black holes, we restrict ourselves to the strong deflection scenario, i.e. to the study of photons having a close approach to the compact lens. We adopt an approximate analytical method, the so-called strong deflection limit \cite{bozza}.

\section{Strong deflection limit}\label{sdl}

\begin{figure}[t!]
\begin{center}
    \includegraphics[scale=0.9,clip=true,angle=0]{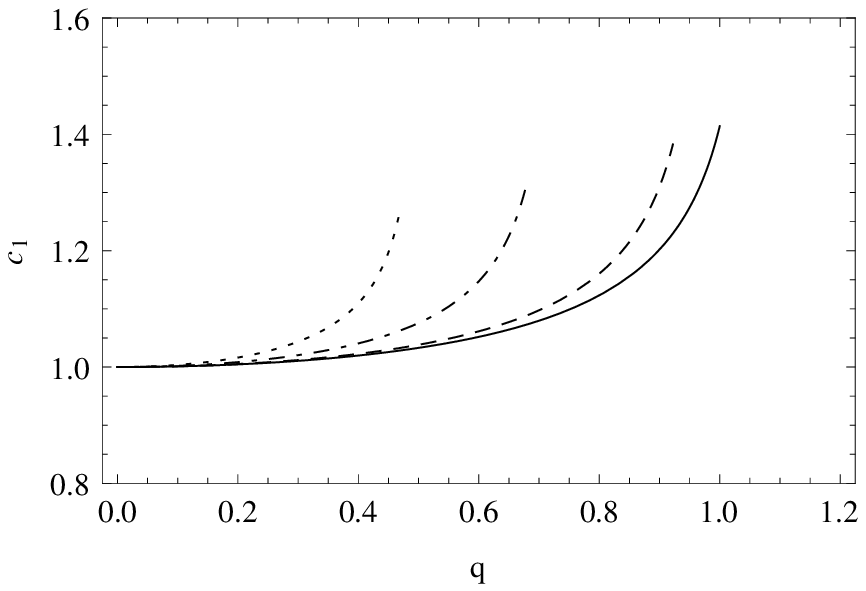}
    \includegraphics[scale=0.9,clip=true,angle=0]{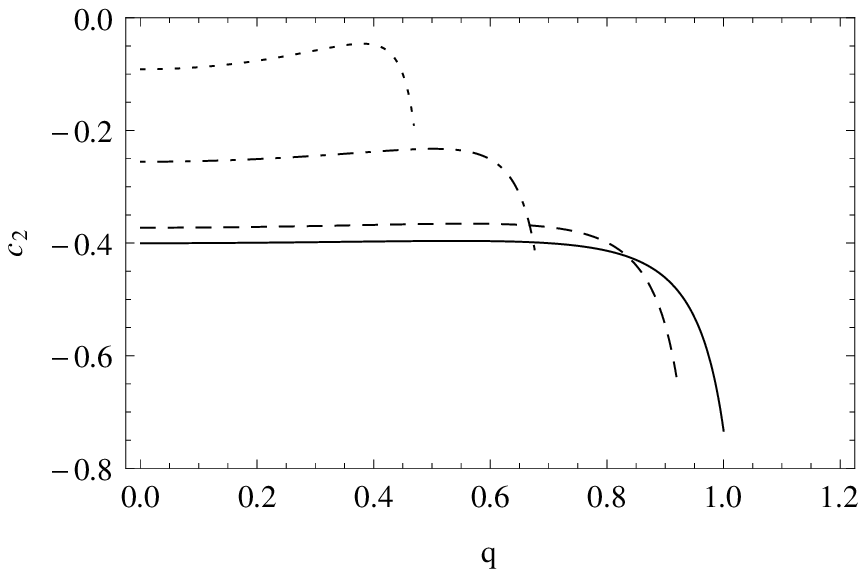}
    \caption{Strong deflection limit coefficients $c_1$ and $c_2$ as functions of the adimensionalized charge $q$ for some representative values of the adimensionalized parameter $\tilde{b}$: $1$ (dashed line), $2$ (dash-dotted line) and $2.5$ (dotted line); the plots corresponding to the Reissner--Nordstr\"om spacetime are also shown (solid line).}
    \label{c1c2}
\end{center}
\end{figure}

When the photons pass close enough to the photon sphere (i.e. when $x_0$ is close to $x_{ps}$), they experiment one or several turns around the black hole  lens before emerging to the observer. The deflection angle is greater than $2\pi$, and two infinite sets of relativistic images are formed, one on each side of the lens. To study this situation, we use that for a spherically symmetric lens the integral (\ref{i0}) can be split \cite{bozza} into a sum of a divergent $I_D(x_0)$ and a regular $I_R(x_0)$ terms:
\begin{equation}
I(x_0)=I_D(x_0)+I_R(x_0),
\label{i0n}
\end{equation}
where
\begin{equation}
I_D(x_0)=\int^{1}_{0}R(0,x_{ps})f_0(z,x_0)dz
\label{id}
\end{equation}
and
\begin{equation}
I_R(x_0)=\int^{1}_{0}[R(z,x_0)f(z,x_0)-R(0,x_{ps})f_0(z,x_0)]dz,
\label{ir}
\end{equation}
with
\begin{equation}
z=\frac{A(x)-A(x_0)}{1-A(x_0)},
\label{z}
\end{equation}
\begin{equation}
R(z,x_0)=\frac{2\sqrt{A(x)B(x)}}{A'(x)C(x)}[1-A(x_0)]\sqrt{C(x_0)},
\label{r}
\end{equation}
\begin{equation}
f(z,x_0)=\frac{1}{\sqrt{A(x_0)-[(1-A(x_0))z+A(x_0)]C(x_0)[C(x)]^{-1}}}.
\label{f}
\end{equation}
After a Taylor expansion of the argument inside the square root in Eq. (\ref{f}) to second order in $z$, one obtains
\begin{equation}
f_0(z,x_0)=\frac{1}{\sqrt{\varphi (x_0) z+\gamma (x_0) z^{2}}},
\label{f0}
\end{equation}
where
\begin{equation}
\varphi (x_0)=\frac{1-A(x_0)}{A'(x_0) C(x_0)}\left[ A(x_0) C'(x_0) - A'(x_0) C(x_0)\right],
\label{varphi}
\end{equation}
\begin{eqnarray}
\gamma (x_0) &=& \frac{\left[ 1-A(x_0)\right] ^{2}}{2[A'(x_0)]^{3} [C(x_0)]^{2}}\left\{ 2 [A'(x_0)]^{2} C(x_0) C'(x_0) - A(x_0) A''(x_0) C(x_0) C'(x_0) \right. \nonumber \\
&& \left. + A(x_0) A'(x_0) \left[ C(x_0) C''(x_0) -2 [C'(x_0)]^{2}\right] \right\} .
\label{gamma}
\end{eqnarray}
The expression $R(z,x_0)$ is regular for all values of $z$ and $x_0$. From Eq. (\ref{varphi}), it is immediate that $\varphi=0$ when $x_0=x_{ps}$, so we have that $f_0\sim1/z$, and the term $I_D$ diverges logarithmically. For all values of $x_0\neq x_{ps}$, we can see that $f_0\sim 1/\sqrt{z}$ and $I_D$ converges. With these definitions, $I_D$ is the term containing the divergence and $I_R$ is regular everywhere, because it has the divergence corresponding to $x_0=x_{ps}$ subtracted. Then, the deflection angle in the strong deflection limit can be written as follows \cite{bozza}
\begin{equation}
\alpha(u)=-c_1\ln\left(\frac{u}{u_{ps}}-1\right)+c_2+O(u-u_{ps}),
\label{alfasdl}
\end{equation}
where
\begin{equation}
u=\sqrt{\frac{C(x_0)}{A(x_0)}},
\label{u}
\end{equation}
is the impact parameter of the photon, and $u_{ps}$ is the impact parameter evaluated at  $x_0=x_{ps}$. The quantities $c_1$ and $c_2$ are the strong deflection limit coefficients, which depend only on the metric functions. In terms of the expressions defined above, they result:
\begin{equation}
c_1=\frac{R(0,x_{ps})}{2\sqrt{\gamma(x_{ps})}}
\label{c1}
\end{equation}
and
\begin{equation}
c_2=-\pi+c_R+c_1\ln \frac{2\gamma(x_{ps})}{A(x_{ps})},
\label{c2}
\end{equation}
with
\begin{equation}
c_R=I_R(x_{ps}).
\label{cr}
\end{equation}
For the charged black hole in scalar--tensor gravity, with metric (\ref{m2}) and $\tilde{b}\neq 0$, we obtain that
\begin{equation}
u_{ps}=\frac{2\tilde{b}^2\sqrt{x^2_{ps}-\tilde{b}^2}}{\sqrt{4\tilde{b}^2\left(\tilde{b}^2-q^2-3x_{ps}\right)+\xi_{ps}\left[-6\tilde{b}^3+2\tilde{b} x_{ps}\left(3x_{ps}+2q^2\right)-q^2\xi_{ps}\left(x^2_{ps}-\tilde{b}^2\right)\right]}},
\end{equation}
\begin{equation}
R(0,x_{ps})= \frac{\sqrt{x^{2}_{ps}-\tilde{b}^2}\left\{4\tilde{b}^2\left(3x_{ps}+q^2\right)+\xi_{ps}\left[6\tilde{b}^3-2\tilde{b}x_{ps}\left(3x_{ps}+2q^2\right)+q^2\xi_{ps}\left(x_{ps}^2-\tilde{b}^2\right)\right]\right\}}{4\tilde{b}^2\left[3\tilde{b}^2-x_{ps}\left(3x_{ps}+q^2\right)\right]+\left(x^{2}_{ps}-\tilde{b}^2\right)\xi_{ps}\left[4\tilde{b}q^2+6\tilde{b} x_{ps}-q^2 x_{ps}\xi_{ps}\right]},
\end{equation}
and
\begin{equation}
\gamma(x_{ps})=-\frac{\Xi\Psi^{2}}{4\tilde{b}\Omega^{3}},
\end{equation}
where
\begin{equation}
\xi_{ps}=\ln\left(\frac{x_{ps}+\tilde{b}}{x_{ps}-\tilde{b}}\right),
\end{equation}
\begin{eqnarray}
\Xi &=& -\tilde{b} q^2\xi^{2}_{ps}(x^{2}_{ps}-\tilde{b}^2) \left\{9 \tilde{b}^2+x_{ps} \left[x_{ps} (4 x_{ps}-45)-18 q^2\right]\right\}-q^4 \xi^{3}_{ps}\left(\tilde{b}^4-6 \tilde{b}^2 x^{2}_{ps}+5x^{4}_{ps}\right) \nonumber
\\
&& +4 \tilde{b}^3 \left\{3 \tilde{b}^4+\tilde{b}^2\left[q^2 (2 x_{ps}-3)+9 x_{ps} (x_{ps}-5)\right]-x_{ps} \left(3 x_{ps}+q^2\right) \left[2q^2+x_{ps} (4 x_{ps}-15)\right]\right\} \nonumber
\\
&&-2 \tilde{b}^2 \xi_{ps}\left\{\tilde{b}^4 \left(2q^2+9\right)-2 \tilde{b}^2 \left[q^4-3 q^2 x_{ps}(x_{ps}-9)-3x^{2}_{ps}(2 x_{ps}-9)\right] \right. \nonumber
\\
&& \left. +x^{2}_{ps} \left[6 q^4-2 q^2x_{ps}(4 x_{ps}-27)-3x^{2}_{ps} (4 x_{ps}-15)\right]\right\},
\end{eqnarray}
\begin{equation}
\Psi=4\tilde{b}^2\left(3x_{ps}+q^2\right)+\xi_{ps}\left[6\tilde{b}^3-2\tilde{b}x_{ps}\left(3x_{ps}+2q^2\right)+q^2\xi_{ps}\left(x^{2}_{ps}-\tilde{b}^2\right)\right],
\end{equation}
and
\begin{equation}
\Omega=4\tilde{b}^2\left[3\tilde{b}^2-x_{ps}\left(3x_{ps}+q^2\right)\right]+\xi_{ps}\left(x^{2}_{ps}-\tilde{b}^2\right)\left[4\tilde{b}q^2+6\tilde{b}x_{ps}-q^2\xi_{ps}x_{ps}\right].
\end{equation}
The integral $c_R$ can be approximated for small values of $\tilde{b}$ and $q$ by its second order Taylor expansion
\begin{equation}
c_R=\ln [36(7-4\sqrt{3})]+\frac{1}{135}(-9+2\sqrt{3})\tilde{b}^2+\frac{2}{9}\left\{ -4+\sqrt{3}+\ln[6(2-\sqrt{3})]\right\}q^2,
\label{crtay} 
\end{equation}
and it can be obtained numerically for arbitrary $\tilde{b}$ and $q$. For the Reissner--Nordstr\"om spacetime, which was studied previously \cite{eiroto,bozza}, the expressions are simpler
\begin{equation}
u_{ps} =\frac{x_{ps}^2}{\sqrt{x_{ps}-q^2}},
\end{equation}
\begin{equation}
R(0,x_{ps})=\frac{2x_{ps}-q^2}{x_{ps}-q^2},
\end{equation}
and
\begin{equation}
\gamma(x_{ps})=\frac{\left(2x_{ps}-q^2\right)^2\left[ 4 q^4 -9 q^2 x_{ps}-(x_{ps}-6) x_{ps}^2\right]}{4 x_{ps}^2 \left( x_{ps}-q^2\right)^3},
\end{equation}
where $x_{ps}=(3+\sqrt{9-8q^2})/2$. The integral $c_R$ in this case can be approximated for small $q$ by taking $\tilde{b}=0$ in Eq. (\ref{crtay}), and calculated numerically for any $q$, using the corresponding metric in Eq. (\ref{cr}). The plots of the strong deflection limit coefficients $c_1$ and $c_2$ as functions of the charge, for different values of the parameter $\tilde{b}$, are shown in Fig. \ref{c1c2}, along with the Reissner--Nordstr\"om ones, for comparison. For a fixed $\tilde{b}$, the coefficient $c_1$ is positive and grows with the charge $q$, when $q=0$ the Schwarzschild value $c_1=1$ is recovered for all $\tilde{b}$; the coefficient $c_2$ is negative and has a slow increase (becomes less negative) with $q$ until it abruptly decreases. For a fixed $q$, the coefficient $c_1$ increases with $\tilde{b}$; the coefficient $c_2$ increases (becomes less negative) with $\tilde{b}$ for a constant (and not very large) $q$.

The positions of the relativistic images are obtained from the lens equation (\ref{pm1}), by replacing the deflection angle (\ref{alfasdl}) in terms of the strong deflection limit coefficients $c_1$ and $c_2$. We consider that the objects are highly aligned, since it is the case for which the lensing effects are more significant. In this approximation, the angles $\beta$ and $\theta$ are small, so the deflection angle for photons passing close to the photon sphere can be written in the form $\alpha=2n\pi+\Delta\alpha_n$, with $n\in\mathbb{N}$, and $0<\Delta\alpha_n\ll 1$. Then, the lens equation (\ref{pm1}) simplifies to
\begin{equation}
\beta =\theta -\frac{d_{ls}}{d_{os}}\Delta \alpha _{n}.
\label{pm2}
\end{equation}
For the photons passing by the other side of the lens we have to replace $\alpha$ by $-\alpha$, i.e. substitute $\Delta \alpha _{n}$ by $-\Delta \alpha _{n}$ in the previous equation. In this small angles approximation, from the lens geometry the impact parameter results in $u=d_{ol}\sin\theta\approx d_{ol}\theta$. Using this relation, inverting Eq. (\ref{alfasdl}), and keeping only the first order term in the Taylor expansion around $\alpha=2n\pi$, the angular position of the $n$th relativistic image is given by
\begin{equation}
\theta _{n}=\theta ^{0}_{n}-\zeta _{n}\Delta \alpha _{n},
\label{pm6}
\end{equation}
where
\begin{equation}
\theta ^{0}_{n}=\frac{u_{ps}}{d_{ol}}\left[ 1+e^{(c_{2}-2n\pi )/c_{1}}
 \right] ,
\label{pm7}
\end{equation}
and
\begin{equation}
\zeta _{n}=\frac{u_{ps}}{c_{1}d_{ol}}e^{(c_{2}-2n\pi )/c_{1}}.
\label{pm8}
\end{equation}
From Eq. (\ref{pm2}) and (\ref{pm6}), we have
$\Delta\alpha_n=(\theta_n-\beta)d_{ol}/d_{ls}$. Then,
\begin{equation}
\theta _{n}=\theta ^{0}_{n}-\frac{\zeta _{n}d_{os}}{d_{ls}}(\theta _{n}-\beta ).
\label{pm10}
\end{equation}
Since $\zeta_n d_{ol}/d_{ls}$ is a small correction to $\theta_n$ because $0<\zeta_n d_{ol}/d_{ls}\ll 1$, the angular positions of the relativistic images finally take the form
\begin{equation}
\theta _{n}=\theta ^{0}_{n}+\frac {\zeta _{n}d_{os}}{d_{ls}}(\beta -\theta ^{0}_{n})
\label{pm14}
\end{equation}
for one set and
\begin{equation}
\theta _{n}=-\theta ^{0}_{n}+\frac {\zeta _{n}d_{os}}{d_{ls}}(\beta +\theta ^{0}_{n})
\label{pm15}
\end{equation}
for the other one.
The magnification of the $n$th relativistic image is given by the quotient of the solid angles subtended by the image and the source
\begin{equation}
\mu=\left|\frac{\sin\beta}{\sin\theta_n}\frac{d\beta}{d\theta_n}\right|^{-1}.
\label{magnification}
\end{equation}
Then, replacing Eq. (\ref{pm14}) in Eq. (\ref{magnification}) and considering small angles, we obtain
\begin{equation}
\mu _{n}=\frac{1}{\beta}\left[ \theta ^{0}_{n}+
\frac {\zeta _{n}d_{os}}{d_{ls}}(\beta - \theta ^{0}_{n})\right]
\frac {\zeta _{n}d_{os}}{d_{ls}},
\label{pm18}
\end{equation}
for both sets of relativistic images. By performing a first order Taylor expansion in $\zeta_n d_{ol}/d_{ls}$, the magnification of the $n$th relativistic image finally results
\begin{equation}
\mu _{n}=\frac{1}{\beta}\frac{\theta ^{0}_{n}\zeta _{n}d_{os}}{d_{ls}}.
\label{pm19}
\end{equation}
By replacing Eqs. (\ref{pm7}) and (\ref{pm8}) in Eq. (\ref{pm19}), we see that the magnifications decrease exponentially with $n$, so the first relativistic image is the brightest one. Unless the lens and the source are highly aligned ($\beta\approx 0$), the magnifications are very faint because the factor $(u_{ps}/d_{ol})^2$ in Eq. (\ref{pm19}) is very small.

\section{Observables}\label{obs}

The direct observation of the supermassive black hole at the Galactic center and also  those present in nearby galaxies is expected to be possible in the next years, when the instruments RADIOASTRON \cite{zakharov,webradio}, Millimetron \cite{johannsen}, Event Horizon Telescope  \cite{eht} and MAXIM \cite{webmaxim} will be operational. RADIOASTRON is a space-based radio telescope, with an angular resolution of about $1-10 \, \mu \mathrm{as}$. The space-based Millimetron mission will have an angular resolution of $0.3$ $\mu$as or less at $0.4$ mm. The Event Horizon telescope  will combine (by using very long baseline interferometry) existing and future radio facilities into a high-sensitivity, high angular resolution telescope. The MAXIM project is a space-based X-ray interferometer with an expected angular resolution of about $0.1 \, \mu \mathrm{as}$. The recent works \cite{sga} discuss the observational aspects of the Galactic supermassive black hole, including the strong deflection of light.

\begin{figure}[t!]
\begin{center}
    \includegraphics[scale=0.9,clip=true,angle=0]{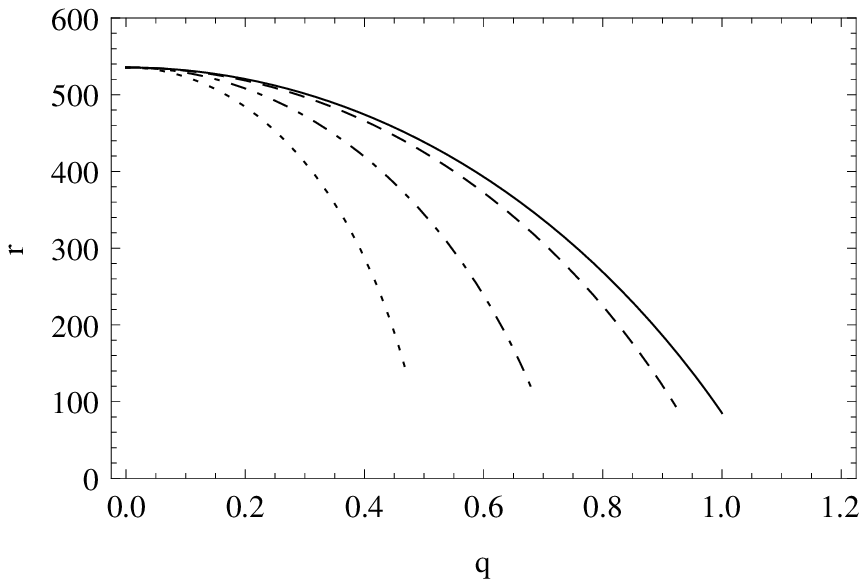}
    \includegraphics[scale=0.9,clip=true,angle=0]{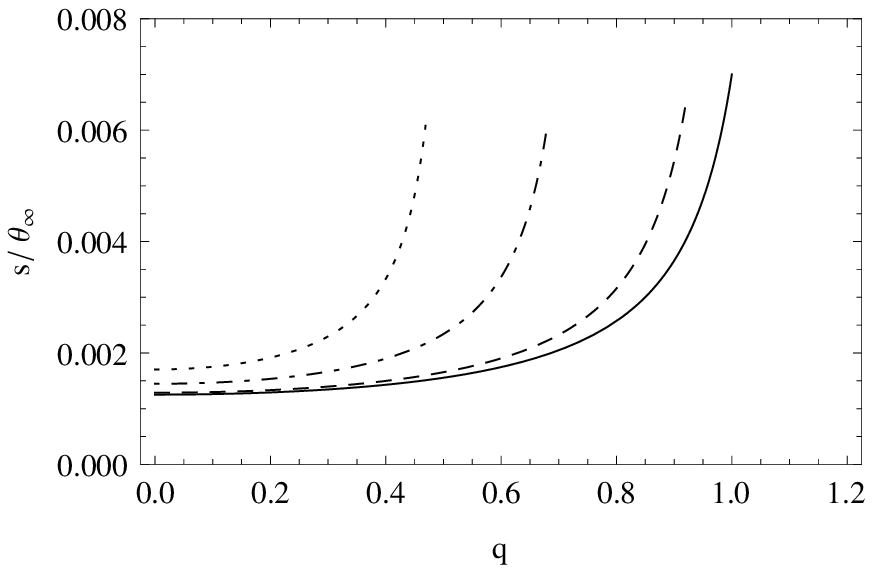}
    \caption{The observables $r$ and $s/\theta_{\infty}$ as functions of the adimensionalized charge $q$ for different values of the adimensionalized parameter $\tilde{b}$: $1$ (dashed line), $2$ (dash-dotted line) and $2.5$ (dotted line); the plots corresponding to the Reissner--Nordstr\"om spacetime are also shown (solid line).}
    \label{observables}
\end{center}
\end{figure}

In order to compare our results with possible future observations, we introduce the quantities \cite{bozza}:
\begin{equation}
\theta_{\infty}=\frac{u_{ps}}{d_{ol}},
\label{ob1}
\end{equation}
\begin{equation}
s=\theta_1-\theta_{\infty},
\label{ob2}
\end{equation}
and
\begin{equation}
r=\frac{\mu_1}{\sum_{n=2}^{\infty}\mu_n}.
\label{ob3}
\end{equation}
The observable $s$ is the angular separation between the first relativistic image (which is the outermost and brightest one) and the limiting value of all the others, which are packed together at $\theta_{\infty}$. The observable $r$ is the quotient between the flux of the first relativistic image and the resulting flux coming from all the others. These observables are defined for the values of $\tilde{b}$ and $q$ for which the relativistic images exist, i.e. when the photon sphere is present. Using expressions (\ref{pm14}) and (\ref{pm19}), it is not difficult to see that these observables, in terms of the strong deflection limit coefficients, are given by
\begin{equation}
s=\theta_{\infty}e^{(c_2-2\pi)/c_1},
\label{ob4}
\end{equation}
and
\begin{equation}
r=e^{2\pi/c_1}.
\label{ob5}
\end{equation}
The plots of $r$ and the quotient $s/\theta_{\infty}$ are shown in Fig. \ref{observables} as functions of $q$ for several values of $\tilde{b}$ and also for the Reissner--Nordstr\"om geometry. We see that for a given value of $\tilde{b}$, the observable $r$ decreases with the charge $q$ from the value $e^{2\pi}$ for $q=0$ to a minimum value in the extremal case, while the quotient $s/\theta_{\infty}$ increases with $q$, reaching a maximum at the extremal value of $q$. For a fixed value of $q$, we have that $r$ decreases with $\tilde{b}$, while $s/\theta_{\infty}$ grows with $\tilde{b}$. In all cases the observables are plot until the charge reaches the largest possible value compatible with the presence of the horizon.

The Galactic center supermassive black hole \cite{guillessen}, has a mass $M=4.31 \times 10^{6}M_{\odot}$ and is situated at a distance from the Earth $D_{ol}=8.33$ kpc. For a numerical example, we can take $D_{os}=2D_{ol}$ as the value of the distance between the observer and the source, an angular position of the source $\beta=0.5 \, \theta_\infty$, and $\tilde{b}=1$. From the equations above, for $ q=0.1$ we obtain that the limiting value of the angular positions of the relativistic images is $\theta_\infty=25.58$ $\mu$as, with the first image separated from it by $s=0.0332$ $\mu$as; the magnification of the first strong deflection image is $\mu_1=6.44\times 10^{-13}$, and the quotient between the flux of the first image and the flux coming from all the others is $r=531.3$. The corresponding values for the Reissner--Nordstr\"om spacetime are $\theta_\infty^{\mathrm{RN}}=26.49$ $\mu$as, $s^{\mathrm{RN}}=0.0334$ $\mu$as, $\mu_1^{\mathrm{RN}}=6.48\times 10^{-13}$, and $r^{\mathrm{RN}}=531.7$. If $ q=0.5$ we find that $\theta_\infty=24.35$ $\mu$as, $s=0.0403$ $\mu$as, $\mu_1=7.54 \times 10^{-13}$, $r=424.9$, and the Reissner--Nordstr\"om values $\theta_\infty^{\mathrm{RN}}=25.37$ $\mu$as, $s^{\mathrm{RN}}=0.0394$ $\mu$as, $\mu_1^{\mathrm{RN}}=7.42 \times 10^{-13}$, and $r^{\mathrm{RN}}=438.3$. The images are highly demagnified, because the value of $\beta $ used in the calculations is not very small compared with $\theta _\infty$. We see that the differences between the results corresponding to the two spacetimes are quite small. The observation of subtle differences coming from the comparison of different black hole models, such as those presented here, will require more advanced future instruments than the ones mentioned above.

\section{Summary}\label{summary}

We have studied the strong deflection lensing effects produced by black holes belonging to a class of charged solutions in scalar--tensor gravity. These spacetimes, characterized by the mass $M$, the charge $Q$, and the parameter $b$ associated to the scalar field, have a horizon surrounded by a photon sphere if the charge is below the extremal one. We have obtained the deflection angle in terms of $b/M$ and $Q/M$ by using the strong deflection limit, from which we have calculated the positions and magnifications of the relativistic images, and also the corresponding observables. We have found that for fixed $b/M$, the limiting value of the image positions $\theta _\infty$ decreases when $Q/M$ increases, the relative separation $s/\theta _\infty$ between the first image and this limiting value grows with $Q/M$, and the relative intensity $r$ of the first image with respect to the others decreases with $Q/M$. For a given value of $Q/M$, we have that $\theta _\infty$ diminishes when $b/M$ grows, $s/\theta _\infty$ increases with $b/M$, and $r$ decreases with $b/M$. In particular, with nonzero $b/M$, for the scalar--tensor black holes we have that $\theta _\infty$ is smaller, $s/\theta _\infty$ is larger, and $r$ is smaller than the corresponding values for the Reissner--Nordstr\"om spacetime with the same value of $Q/M$. The differences between the results for these geometries are very small, not detectable with current or near future facilities.

\section*{Acknowledgments}

This work has been supported by CONICET and  University of Buenos Aires.

\end{document}